\documentclass[twoside,11pt]{article}  

\usepackage{subfigure}
\usepackage{enumerate}
\usepackage{bm}
\usepackage{amsmath}
\usepackage{amsthm}
\usepackage{amssymb}
\usepackage{wrapfig}
\usepackage{array}
\usepackage{color}
\usepackage{algorithmic}
\usepackage{graphics}
\usepackage{natbib}

\usepackage{tikz}
\usetikzlibrary{arrows,arrows.meta}
\usetikzlibrary{decorations.markings}
\usetikzlibrary{decorations.pathmorphing}
\usetikzlibrary{positioning,shapes,backgrounds}
\usetikzlibrary{calc,intersections,patterns,fit,automata,snakes}
\usepackage{tikzsymbols}

\newcommand{\remove}[1]{}

\definecolor{skipcolor}{rgb}{1, .92, .92}
\definecolor{rejoincolor}{rgb}{.9, 1, 0.9}
\definecolor{frameskip}{rgb}{0.95, 0, 0}
\definecolor{framerejoin}{rgb}{0, .75, 0}

\definecolor{gold}{rgb}{1,0.553,0}
\definecolor{lightbrown}{rgb}{0.9305,0.86275,0.8}
\definecolor{fillcopper}{rgb}{0.722,0.451,0.2}
\definecolor{fillsilver}{rgb}{0.753,0.753,0.753}
\definecolor{fillgray}{rgb}{0.4,0.4,0.4}
\definecolor{filllightgray}{rgb}{0.65,0.65,0.65}
\definecolor{fillLightgray}{rgb}{0.85,0.85,0.85}
\definecolor{fillred}{rgb}{1,0.2,0.2}
\definecolor{filllightred}{rgb}{1,0.6,0.6}
\definecolor{fillblue}{rgb}{0.2,0.35,1}
\definecolor{filllightblue}{rgb}{0.85,0.85,1}
\definecolor{fillgreen}{rgb}{0.2,0.7,0.2}
\definecolor{filllightgreen}{rgb}{0.9,1,0.9}
\definecolor{fillgreenbox}{rgb}{0.9,1,0.9}
\definecolor{fillpurple}{rgb}{0.7,0.4,0.74}

\def\math#1{$#1$}

\def\mand#1{$$#1$$}

\def\mymathskip{4.5pt}

\def\mandc#1{\mand{\abovedisplayskip=\mymathskip plus 1pt minus 1pt%
\abovedisplayshortskip=0pt plus 1pt minus 1pt%
\belowdisplayskip=\mymathskip plus 1pt minus 1pt%
\belowdisplayshortskip=0pt plus 1pt minus 1pt%
#1}}

\def\mldc#1{\mld{\abovedisplayskip=\mymathskip plus 1pt minus 1pt%
\abovedisplayshortskip=0pt plus 1pt minus 1pt%
\belowdisplayskip=\mymathskip plus 1pt minus 1pt%
\belowdisplayshortskip=0pt plus 1pt minus 1pt%
#1}}

\def\eqarc#1{\eqar{\abovedisplayskip=\mymathskip plus 1pt minus 1pt%
\abovedisplayshortskip=0pt plus 1pt minus 1pt%
\belowdisplayskip=\mymathskip plus 1pt minus 1pt%
\belowdisplayshortskip=0pt plus 1pt minus 1pt%
#1}}

\def\mld#1{\begin{equation}
#1
\end{equation}}

\def\eqar#1{\begin{eqnarray}
#1
\end{eqnarray}}

\def\frac#1#2{{#1\over #2}}

\DeclareSymbolFont{AMSb}{U}{msb}{m}{n}

\def\cl#1{{\cal #1}}

\def\ceil#1{
\mathchoice
{\left\lceil\,#1\,\right\rceil}
{\big\lceil\,#1\,\big\rceil}
{\lceil\,#1\,\rceil}
{\lceil\,#1\,\rceil}
}

\def\r#1{{\eqref{#1}}}

\providecommand{\red}[1]{{\color{fillred} #1}}
\providecommand{\blue}[1]{{\color{fillblue} #1}}

\newcounter{rmnum}
\def\RN#1{\setcounter{rmnum}{#1}\uppercase\expandafter{\romannumeral\value{rmnum}}}
\def\rn#1{\setcounter{rmnum}{#1}\expandafter{\romannumeral\value{rmnum}}}

\remove{

}

\definecolor{shadecolor}{gray}{.85}%
\definecolor{tintedcolor}{gray}{.8}%

{\makeatletter\gdef\reallynopagebreak{\par\nopagebreak\@nobreaktrue}}

\providecommand\remove[1]{}

\DeclareSymbolFont{extraup}{U}{zavm}{m}{n}
\DeclareMathSymbol{\varheart}{\mathalpha}{extraup}{86}
\DeclareMathSymbol{\vardiamond}{\mathalpha}{extraup}{87}

\newcommand{\myslimfbox}[3][white]{%
\begin{center}%
\vspace*{-5pt}%
\begin{tikzpicture}%
\node[draw,inner sep=5pt,rounded corners=5pt,line width=1pt,fill=#1] at (0,0)
{\parbox{#2}{\centering #3}};
\end{tikzpicture}%
\vspace*{-5pt}%
\end{center}%
}

\newcommand{\dieface}[1]{%
\begin{tikzpicture}[baseline=-5pt,line width=2pt]
\begin{scope}
\coordinate(center)at(0,0);
\ifnum#1=1
\node[circle,draw,fill,inner sep=1pt] at(center){};
\fi
\ifnum#1=2
\node[circle,draw,fill,inner sep=1pt] at($(center)+(-0.2,0)$){};
\node[circle,draw,fill,inner sep=1pt] at($(center)+(0.2,0)$){};
\fi
\ifnum#1=3
\node[circle,draw,fill,inner sep=1pt] at($(center)+(-0.2,-0.2)$){};
\node[circle,draw,fill,inner sep=1pt] at($(center)+(0,0.2)$){};
\node[circle,draw,fill,inner sep=1pt] at($(center)+(0.2,-0.2)$){};
\fi
\ifnum#1=4
\node[circle,draw,fill,inner sep=1pt] at($(center)+(-0.2,0.2)$){};
\node[circle,draw,fill,inner sep=1pt] at($(center)+(0.2,0.2)$){};
\node[circle,draw,fill,inner sep=1pt] at($(center)+(-0.2,-0.2)$){};
\node[circle,draw,fill,inner sep=1pt] at($(center)+(0.2,-0.2)$){};
\fi
\ifnum#1=5
\node[circle,draw,fill,inner sep=1pt] at($(center)+(-0.25,0.25)$){};
\node[circle,draw,fill,inner sep=1pt] at($(center)+(0.25,0.25)$){};
\node[circle,draw,fill,inner sep=1pt] at($(center)+(0,0)$){};
\node[circle,draw,fill,inner sep=1pt] at($(center)+(-0.25,-0.25)$){};
\node[circle,draw,fill,inner sep=1pt] at($(center)+(0.25,-0.25)$){};
\fi
\ifnum#1=6
\node[circle,draw,fill,inner sep=1pt] at($(center)+(-0.2,0.25)$){};
\node[circle,draw,fill,inner sep=1pt] at($(center)+(0.2,0.25)$){};
\node[circle,draw,fill,inner sep=1pt] at($(center)+(-0.2,0)$){};
\node[circle,draw,fill,inner sep=1pt] at($(center)+(0.2,0)$){};
\node[circle,draw,fill,inner sep=1pt] at($(center)+(-0.2,-0.25)$){};
\node[circle,draw,fill,inner sep=1pt] at($(center)+(0.2,-0.25)$){};
\fi
\ifnum#1=7
\node[circle,draw,fill,inner sep=1pt] at($(center)+(-0.15,0.25)$){};
\node[circle,draw,fill,inner sep=1pt] at($(center)+(0.15,0.25)$){};
\node[circle,draw,fill,inner sep=1pt] at($(center)+(-0.25,0)$){};
\node[circle,draw,fill,inner sep=1pt] at($(center)+(0,0)$){};
\node[circle,draw,fill,inner sep=1pt] at($(center)+(0.25,0)$){};
\node[circle,draw,fill,inner sep=1pt] at($(center)+(-0.15,-0.25)$){};
\node[circle,draw,fill,inner sep=1pt] at($(center)+(0.15,-0.25)$){};
\fi
\ifnum#1=8
\node[circle,draw,fill,inner sep=1pt] at($(center)+(-0.25,0.25)$){};
\node[circle,draw,fill,inner sep=1pt] at($(center)+(0.25,0.25)$){};
\node[circle,draw,fill,inner sep=1pt] at($(center)+(-0.25,0)$){};
\node[circle,draw,fill,inner sep=1pt] at($(center)+(0,0.15)$){};
\node[circle,draw,fill,inner sep=1pt] at($(center)+(0,-0.15)$){};
\node[circle,draw,fill,inner sep=1pt] at($(center)+(0.25,0)$){};
\node[circle,draw,fill,inner sep=1pt] at($(center)+(-0.25,-0.25)$){};
\node[circle,draw,fill,inner sep=1pt] at($(center)+(0.25,-0.25)$){};
\fi
\ifnum#1=9
\node[circle,draw,fill,inner sep=1pt] at($(center)+(-0.25,0.25)$){};
\node[circle,draw,fill,inner sep=1pt] at($(center)+(0.25,0.25)$){};
\node[circle,draw,fill,inner sep=1pt] at($(center)+(-0.25,0)$){};
\node[circle,draw,fill,inner sep=1pt] at($(center)+(0,0.25)$){};
\node[circle,draw,fill,inner sep=1pt] at($(center)+(0,0)$){};
\node[circle,draw,fill,inner sep=1pt] at($(center)+(0,-0.25)$){};
\node[circle,draw,fill,inner sep=1pt] at($(center)+(0.25,0)$){};
\node[circle,draw,fill,inner sep=1pt] at($(center)+(-0.25,-0.25)$){};
\node[circle,draw,fill,inner sep=1pt] at($(center)+(0.25,-0.25)$){};
\fi
\draw[rounded corners=8pt]($(center)+(-0.45,-0.45)$)rectangle($(center)+(0.45,0.45)$);
\end{scope}
\end{tikzpicture}
}
\newsavebox{\done}\begin{lrbox}{\done}\dieface{1}\end{lrbox}
\newsavebox{\dtwo}\begin{lrbox}{\dtwo}\dieface{2}\end{lrbox}
\newsavebox{\dthree}\begin{lrbox}{\dthree}\dieface{3}\end{lrbox}
\newsavebox{\dfour}\begin{lrbox}{\dfour}\dieface{4}\end{lrbox}
\newsavebox{\dfive}\begin{lrbox}{\dfive}\dieface{5}\end{lrbox}
\newsavebox{\dsix}\begin{lrbox}{\dsix}\dieface{6}\end{lrbox}
\newsavebox{\dseven}\begin{lrbox}{\dseven}\dieface{7}\end{lrbox}
\newsavebox{\deight}\begin{lrbox}{\deight}\dieface{8}\end{lrbox}
\newsavebox{\dnine}\begin{lrbox}{\dnine}\dieface{9}\end{lrbox}

\newcommand{\domsmall}[1]{%
\begin{tikzpicture}[baseline=-5pt,line width=3pt]
\begin{scope}
\coordinate(center)at(0,0);
\ifnum#1=1
\node[circle,draw,fill,inner sep=1pt] at(center){};
\fi
\ifnum#1=2
\node[circle,draw,fill,inner sep=1pt] at($(center)+(-0.2,0)$){};
\node[circle,draw,fill,inner sep=1pt] at($(center)+(0.2,0)$){};
\fi
\ifnum#1=3
\node[circle,draw,fill,inner sep=1pt] at($(center)+(-0.2,-0.2)$){};
\node[circle,draw,fill,inner sep=1pt] at($(center)+(0,0.2)$){};
\node[circle,draw,fill,inner sep=1pt] at($(center)+(0.2,-0.2)$){};
\fi
\ifnum#1=4
\node[circle,draw,fill,inner sep=1pt] at($(center)+(-0.2,0.2)$){};
\node[circle,draw,fill,inner sep=1pt] at($(center)+(0.2,0.2)$){};
\node[circle,draw,fill,inner sep=1pt] at($(center)+(-0.2,-0.2)$){};
\node[circle,draw,fill,inner sep=1pt] at($(center)+(0.2,-0.2)$){};
\fi
\ifnum#1=5
\node[circle,draw,fill,inner sep=1pt] at($(center)+(-0.25,0.25)$){};
\node[circle,draw,fill,inner sep=1pt] at($(center)+(0.25,0.25)$){};
\node[circle,draw,fill,inner sep=1pt] at($(center)+(0,0)$){};
\node[circle,draw,fill,inner sep=1pt] at($(center)+(-0.25,-0.25)$){};
\node[circle,draw,fill,inner sep=1pt] at($(center)+(0.25,-0.25)$){};
\fi
\ifnum#1=6
\node[circle,draw,fill,inner sep=1pt] at($(center)+(-0.2,0.25)$){};
\node[circle,draw,fill,inner sep=1pt] at($(center)+(0.2,0.25)$){};
\node[circle,draw,fill,inner sep=1pt] at($(center)+(-0.2,0)$){};
\node[circle,draw,fill,inner sep=1pt] at($(center)+(0.2,0)$){};
\node[circle,draw,fill,inner sep=1pt] at($(center)+(-0.2,-0.25)$){};
\node[circle,draw,fill,inner sep=1pt] at($(center)+(0.2,-0.25)$){};
\fi
\ifnum#1=7
\node[circle,draw,fill,inner sep=1pt] at($(center)+(-0.15,0.25)$){};
\node[circle,draw,fill,inner sep=1pt] at($(center)+(0.15,0.25)$){};
\node[circle,draw,fill,inner sep=1pt] at($(center)+(-0.25,0)$){};
\node[circle,draw,fill,inner sep=1pt] at($(center)+(0,0)$){};
\node[circle,draw,fill,inner sep=1pt] at($(center)+(0.25,0)$){};
\node[circle,draw,fill,inner sep=1pt] at($(center)+(-0.15,-0.25)$){};
\node[circle,draw,fill,inner sep=1pt] at($(center)+(0.15,-0.25)$){};
\fi
\ifnum#1=8
\node[circle,draw,fill,inner sep=1pt] at($(center)+(-0.25,0.25)$){};
\node[circle,draw,fill,inner sep=1pt] at($(center)+(0.25,0.25)$){};
\node[circle,draw,fill,inner sep=1pt] at($(center)+(-0.25,0)$){};
\node[circle,draw,fill,inner sep=1pt] at($(center)+(0,0.15)$){};
\node[circle,draw,fill,inner sep=1pt] at($(center)+(0,-0.15)$){};
\node[circle,draw,fill,inner sep=1pt] at($(center)+(0.25,0)$){};
\node[circle,draw,fill,inner sep=1pt] at($(center)+(-0.25,-0.25)$){};
\node[circle,draw,fill,inner sep=1pt] at($(center)+(0.25,-0.25)$){};
\fi
\ifnum#1=9
\node[circle,draw,fill,inner sep=1pt] at($(center)+(-0.25,0.25)$){};
\node[circle,draw,fill,inner sep=1pt] at($(center)+(0.25,0.25)$){};
\node[circle,draw,fill,inner sep=1pt] at($(center)+(-0.25,0)$){};
\node[circle,draw,fill,inner sep=1pt] at($(center)+(0,0.25)$){};
\node[circle,draw,fill,inner sep=1pt] at($(center)+(0,0)$){};
\node[circle,draw,fill,inner sep=1pt] at($(center)+(0,-0.25)$){};
\node[circle,draw,fill,inner sep=1pt] at($(center)+(0.25,0)$){};
\node[circle,draw,fill,inner sep=1pt] at($(center)+(-0.25,-0.25)$){};
\node[circle,draw,fill,inner sep=1pt] at($(center)+(0.25,-0.25)$){};
\fi
\draw($(center)+(-0.45,-0.45)$)rectangle($(center)+(0.45,0.45)$);
\end{scope}
\end{tikzpicture}
}

\newsavebox\mandown
\begin{lrbox}{\mandown}
\begin{tikzpicture}[line width=2pt]
\shade[ball color=fillgreen]
(-0.7,-0.5) 
to [bend right=20](-1.3,-2.5) 
to[bend right=10] (-0.9,-2.48)
to [bend right=5](-0.6,-4.45)
to [bend right=5](0.6,-4.35)
to [bend right=5](1.8,-4.2)
to [out=88,in=-89](1.4,-1.2)
to (1.5,-2)
to [bend right=3](1.9,-1.9)
to [bend right=5](0.9,0.7)
--cycle;
\path
(-0.7,-0.5)edge[bend right=20](-1.3,-2.5)
(-1.3,-2.5)edge[bend right=10](-0.3,-2.4)
(-0.3,-2.4)edge[bend left=10](-0.3,-1.5)
(-0.9,-2.48)edge[bend right=5](-0.6,-4.45)
(-0.6,-4.45)edge[bend right=5](0.6,-4.35)
(0.6,-4.3)edge[bend right=5](1.8,-4.2)
(0.6,-4.3)edge[bend right=0](0.58,-3.2)
(0.4,-3.1)edge[bend left=5](0.76,-3.1)
(1.8,-4.1)edge[out=88,in=-89](1.4,-1.2)
(1.5,-2)edge[bend right=3](1.9,-1.9)
(1.9,-1.9)edge[bend right=5](0.9,0.7)
;
\shade[ball color=fillred,label=E1,draw,line width=1pt](0,0.5)ellipse(1.75cm and 1.75cm);
\draw[fill,rotate=10](-0.9,-0.4)ellipse(1mm and 2mm);
\draw[fill,rotate=-10](0.9,-0.4)ellipse(1mm and 2mm);
\end{tikzpicture}
\end{lrbox}

\newsavebox\nodebox
\begin{lrbox}{\nodebox}
\begin{tikzpicture}
\node[circle,fill=fillred,inner sep=2pt] at (0,0){};
\end{tikzpicture}
\end{lrbox}

\newsavebox{\DMC}
\begin{lrbox}{\DMC}
\begin{tikzpicture}[line width=1.5pt]
\node[draw,rounded corners=10pt,minimum height=1.6cm,minimum width=1.7cm,fill=fillgray,double,double distance=3pt](B)at(0,0){};
\node[inner sep=1pt](M)at(-0.175,0){\scalebox{2}[3]{\math{\bm{\cl{M}}}}};
\node[inner sep=1pt](sq)at($(M.east|-M.north)+(0.1,-0.1)$)
{\scalebox{1.5}{\bf\textsc{2}}};
\end{tikzpicture}
\end{lrbox}

\newcommand{\TruthTable}[4]{
\begin{tikzpicture}[x=0.65cm,y=0.375cm,baseline=-3pt]
\pgfmathtruncatemacro\R{2^(#1)};
\foreach\x[count=\i]in{#2}{
\node[](v\i)at(0.7*\i,0){\math{\x}};
}
\node[anchor=west](prop)at($(v#1)+(0.5,0)$){#3};
\draw($(v1.west)+(0,-0.5)$)--($(prop.east)+(0,-0.5)$);
\draw($0.5*(v#1.east)+0.5*(prop.west)+(0,0.5)$)--($0.5*(v#1.east)+0.5*(prop.west)+(0,-\R-0.5)$);
\foreach\t[count=\i] in{#4}{
\node[](tv\i)at($(prop.south)+(0,-\i+0.5)$){\t};
}
\foreach\c in {1,...,#1}{
\foreach\r in {1,...,\R}{
\pgfmathtruncatemacro\s{2^(#1-\c)};
\pgfmathtruncatemacro\a{mod(ceil(\r/\s),2)};
\def\tval{\F};
\ifnum\a<1
\def\tval{\T};
\fi
\node[]at(v\c|-tv\r){\tval};
}}
\end{tikzpicture}
}

\newcommand{\TruthTableTwo}[6]{
\begin{tikzpicture}[x=0.65cm,y=0.375cm,baseline=-3pt]
\pgfmathtruncatemacro\R{2^(#1)};
\foreach\x[count=\i]in{#2}{
\node[](v\i)at(0.7*\i,0){\math{\x}};
}
\node[anchor=west](prop1)at($(v#1)+(0.5,0)$){#3};
\node[anchor=west](prop2)at($(prop1.east)+(0.5,0)$){#5};
\draw($(v1.west)+(0,-0.5)$)--($(prop2.east)+(0,-0.5)$);
\draw($0.5*(v#1.east)+0.5*(prop1.west)+(0,0.5)$)--($0.5*(v#1.east)+0.5*(prop1.west)+(0,-\R-0.5)$);
\foreach\t[count=\i] in{#4}{
\node[](tv\i)at($(prop1.south)+(0,-\i+0.5)$){\t};
}
\foreach\t[count=\i] in{#6}{
\node[](tv\i)at($(prop2.south)+(0,-\i+0.5)$){\t};
}
\foreach\c in {1,...,#1}{
\foreach\r in {1,...,\R}{
\pgfmathtruncatemacro\s{2^(#1-\c)};
\pgfmathtruncatemacro\a{mod(ceil(\r/\s),2)};
\def\tval{\F};
\ifnum\a<1
\def\tval{\T};
\fi
\node[]at(v\c|-tv\r){\tval};
}}
\end{tikzpicture}
}
\newcommand{\TruthTableThree}[8]{
\begin{tikzpicture}[x=0.65cm,y=0.375cm,baseline=-3pt]
\pgfmathtruncatemacro\R{2^(#1)};
\foreach\x[count=\i]in{#2}{
\node[](v\i)at(0.7*\i,0){\math{\x}};
}
\node[anchor=west](prop1)at($(v#1)+(0.5,0)$){#3};
\node[anchor=west](prop2)at($(prop1.east)+(0.5,0)$){#5};
\node[anchor=west](prop3)at($(prop2.east)+(0.5,0)$){#7};
\draw($(v1.west)+(0,-0.5)$)--($(prop3.east)+(0,-0.5)$);
\draw($0.5*(v#1.east)+0.5*(prop1.west)+(0,0.5)$)--($0.5*(v#1.east)+0.5*(prop1.west)+(0,-\R-0.5)$);
\foreach\t[count=\i] in{#4}{
\node[](tv\i)at($(prop1.south)+(0,-\i+0.5)$){\t};
}
\foreach\t[count=\i] in{#6}{
\node[](tv\i)at($(prop2.south)+(0,-\i+0.5)$){\t};
}
\foreach\t[count=\i] in{#8}{
\node[](tv\i)at($(prop3.south)+(0,-\i+0.5)$){\t};
}
\foreach\c in {1,...,#1}{
\foreach\r in {1,...,\R}{
\pgfmathtruncatemacro\s{2^(#1-\c)};
\pgfmathtruncatemacro\a{mod(ceil(\r/\s),2)};
\def\tval{\F};
\ifnum\a<1
\def\tval{\T};
\fi
\node[]at(v\c|-tv\r){\tval};
}}
\end{tikzpicture}
}

\newcommand{\venntwo}[2]{
\def\firstcircle{(0,0) circle (1.5cm)}
\def\secondcircle{(0:2cm) circle (1.5cm)}
\def\universal{(-2,-2) rectangle (4,2)}
\foreach\x[count=\i] in {#2}{
\ifnum\i=1
\fill[fill=\x] \universal;
\fi
\ifnum\i=2
\fill[fill=\x] \firstcircle;
\fi
\ifnum\i=3
\fill[fill=\x] \secondcircle;
\fi
\ifnum\i=4
\begin{scope}
\clip\firstcircle;
\fill[fill=\x] \secondcircle;
\end{scope}
\fi
}
\foreach\x[count=\i] in {#1}{
\ifnum\i=1
\draw\firstcircle node[scale=0.9,left=-1pt]{\x};
\fi
\ifnum\i=2
\draw\secondcircle node[scale=0.9,right=-1pt]{\x};
\fi
}
\draw\universal;
}

\newcommand{\vennthree}[2]{
\def\firstcircle{(0,0) circle (1.5cm)}
\def\secondcircle{(60:2cm) circle (1.5cm)}
\def\thirdcircle{(0:2cm) circle (1.5cm)}
\def\universal{(-2,-2) rectangle (4,3.7321)}
\foreach\x[count=\i] in {#2}{
\ifnum\i=1
\fill[fill=\x] \universal;
\fi
\ifnum\i=2
\fill[fill=\x] \firstcircle;
\fi
\ifnum\i=3
\fill[fill=\x] \secondcircle;
\fi
\ifnum\i=4
\fill[fill=\x] \thirdcircle;
\fi
\ifnum\i=5
\begin{scope}
\clip\firstcircle;
\fill[fill=\x] \secondcircle;
\end{scope}
\fi
\ifnum\i=6
\begin{scope}
\clip\firstcircle;
\fill[fill=\x] \thirdcircle;
\end{scope}
\fi
\ifnum\i=7
\begin{scope}
\clip\secondcircle;
\fill[fill=\x] \thirdcircle;
\end{scope}
\fi
\ifnum\i=8
\begin{scope}
\clip\firstcircle;
\clip\secondcircle;
\fill[fill=\x] \thirdcircle;
\end{scope}
\fi
}
\foreach\x[count=\i] in {#1}{
\ifnum\i=1
\draw\firstcircle node[scale=0.9,left=-1pt]{\x};
\fi
\ifnum\i=2
\draw\secondcircle node[scale=0.9,above=-1pt]{\x};
\fi
\ifnum\i=3
\draw\thirdcircle node[scale=0.9,right=-1pt]{\x};
\fi
}
\draw\universal;
}

\DeclareMathSymbol{\Prob}{\mathbin}{AMSb}{"50}
\DeclareMathSymbol{\Exp}{\mathbin}{AMSb}{"45}

\title{Machine Learning the Phenomenology of COVID-19 From Early Infection Dynamics}
\author{Malik Magdon-Ismail\\
magdon@cs.rpi.edu\\
Computer Science Department\\
Rensselaer Ploytechnic Institute\\
110 8th Street, Troy, NY 12180, USA
}

\begin{document}

\maketitle

\begin{abstract}%
  \noindent
   We present a robust data-driven machine learning analysis of the
COVID-19 pandemic from its
  \emph{early} infection dynamics, specifically infection counts over time.
  The goal is to extract
  actionable public health insights.
  These insights include the infectious force, the rate of a mild infection
  becoming serious, estimates for asymtomatic infections and
  predictions of new infections over time. 
  We focus on 
  USA data starting from the first confirmed
  infection on January 20 2020.
  Our methods reveal significant asymptomatic
  (hidden) infection, a lag of about 10 days, and we quantitatively
  confirm that the infectious force is strong with about a 0.14\%
  transition from mild to serious infection.
  Our methods are efficient,
  robust and general, being
  agnostic to the specific
  virus and applicable to different populations or cohorts.

\end{abstract}

\section{Introduction}
\label{section:intro}

As of March 1 2020, there was still much public debate on properties
of the COVID-19 pandemic
(see the CNN article, \cite{CNN2020}). For example, is
asymptomatic spread of COVID-19 a major driver of the pandemic?
There was no clear unimodal view, highlighting the
need for robust tools to generate actionable quantitative
intelligence on the nature
of a pandemic from \emph{early} and minimal data.
One approach is scenario analysis.
Recently, \cite{Chinazzi2020} used
the Global Epidemic and Mobility Model (GLEAM)
to perform infection scenario analyses on China COVID-19
data, using
a networked meta-population model based on transportation hubs.
A similar model for the US was reported in \cite{TIME2020} where the
web-app predicted from 150,000 to 1.4 million infected cases by April 30,
depending on the intervention level.
Such scenario analysis requires user input such as infection
sites and contagion-properties. However,
a majority of infection sites may be hidden, especially if asymptomatic
transmission is significant. Further, the contagion parameters are
\emph{unknown} and 
must be deduced, perhaps using domain expertise.

Data driven approaches are powerful. A long range regression analysis of
COVID-19 out to 2025 on US data using immunological,
epidemiological and seasonal
effects is given in \cite{Kissler2020},
which predicts  recurrent outbreaks.
We also follow a data-driven machine learning
approach to understand early dynamics of COVID-19 on the first
54 days of US confirmed infection reports (downloadable from
the European Center For Disease Control).
We address the challenge of real-time data-intelligence from early data.
Our approach
is simple, requires minimal data or human input and generates
actionable insights.
For example, is asymptomatic spread significant?
Our data-driven analysis says yes, emphatically. We even give quantitative
estimates
for the number of asymptomatic infections.

Early data
is both a curse and a blessing.
The curse is that ``early'' implies not
much information, so quantitative models must be simple and robust
to be identifiable from the data.
The blessing is that early data is
a peek at the pandemic as it roams free,
unchecked by public health protocols, for
to learn the true intentions of the lion, you must follow the
beast on the savanna, not in the zoo.
As we will see, much can be learned from early data and
these insights early in the game,
can be crucial to public health governance.
\begin{figure}[t]
  \centering
  \begin{tikzpicture}[>=latex,line width=1pt]
    \node[inner sep=0pt](F)at(0,0){\includegraphics[width=0.7\textwidth,trim={1cm 6.7cm 1.35cm 6.9cm},clip]{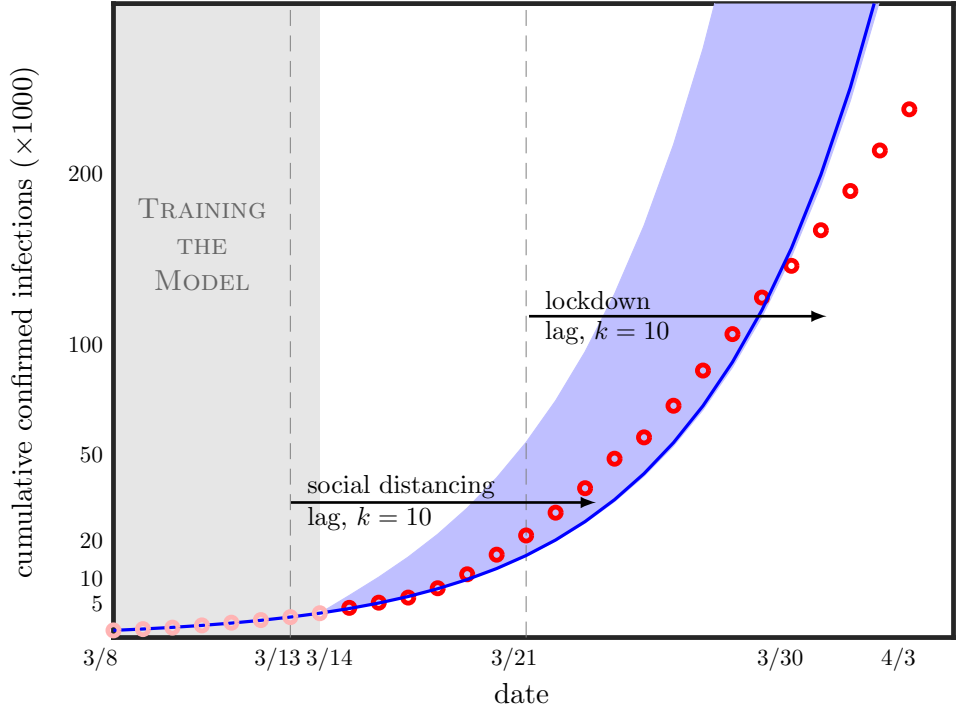}};
    \coordinate(TL)at(F.west|-F.north);
    \coordinate(BL)at(F.west|-F.south);
    \coordinate(TR)at(F.east|-F.north);
    \coordinate(BR)at(F.east|-F.south);    
    \def\Tbeg{47}
    \def\Tmid{53}
    \def\Tlock{61}
    \def\Tend{75.5}
    \def\Tfinal{74}
    \def\datefinal{4/3}
    \def\ymin{1}
    \def\ymax{3826.0}
    \def\pow{0.65}

    \foreach\y/\x[count=\i]in{47/{3/8},\Tmid/{3/13},54.75/{3/14},\Tlock/{3/21},70/{3/30},\Tfinal/{\datefinal}}{
      \pgfmathsetmacro\lam{1};
      \pgfmathsetmacro\lambar{1-\lam};
      \pgfmathsetmacro\z{1-(\y-\Tbeg+0.5)/(\Tend-\Tbeg+1)}
      \pgfmathsetmacro\zbar{1-\z};
      \node[scale=0.8,inner sep=0pt,anchor=north](x\i)at($\z*\lam*(BL)+\z*\lambar*(TL)+\zbar*\lam*(BR)+\zbar*\lambar*(TR)$){\x};
    }
    \foreach\y/\ty[count=\i]in{50/5,100/10,200/20,500/50,1000/100,2000/200}{
      \pgfmathsetmacro\lam{0.97};
      \pgfmathsetmacro\lambar{1-\lam};
      \pgfmathsetmacro\z{1-(100^\pow*\y^(\pow)-\ymin+6)/(\ymax-\ymin+12)}
      \pgfmathsetmacro\zbar{1-\z};
      \node[scale=0.8,anchor=east](x\i)at($\z*\lam*(BL)+\z*\lambar*(BR)+\zbar*\lam*(TL)+\zbar*\lambar*(TR)$){\ty};
    }
    \pgfmathsetmacro\xlam{(\Tlock.5-\Tbeg+0.4)/(\Tend-\Tbeg+0.8)};
    \pgfmathsetmacro\xlambar{1-\xlam};
    \coordinate(xlock)at($\xlambar*(BL)+\xlam*(BR)$);
    \pgfmathsetmacro\xlam{(\Tlock.5+10-\Tbeg+0.4)/(\Tend-\Tbeg+0.8)};
    \pgfmathsetmacro\xlambar{1-\xlam};
    \coordinate(xlocklag)at($\xlambar*(BL)+\xlam*(BR)$);
    \pgfmathsetmacro\ylam{(100^(\pow)*1150^(\pow)-\ymin+6)/(\ymax-\ymin+12)};
    \pgfmathsetmacro\ylambar{1-\ylam};
    \coordinate(ylock)at($\ylambar*(BL)+\ylam*(TL)$);
    \pgfmathsetmacro\xlam{(\Tmid.5-\Tbeg+0.4)/(\Tend-\Tbeg+0.8)};
    \pgfmathsetmacro\xlambar{1-\xlam};
    \coordinate(xlast)at($\xlambar*(BL)+\xlam*(BR)$);
    \pgfmathsetmacro\ylam{(100^(\pow)*320^(\pow)-\ymin+6)/(\ymax-\ymin+12)};
    \pgfmathsetmacro\ylambar{1-\ylam};
    \coordinate(ylast)at($\ylambar*(BL)+\ylam*(TL)$);
    \pgfmathsetmacro\xlam{(\Tmid.5+10.25-\Tbeg+0.4)/(\Tend-\Tbeg+0.8)};
    \pgfmathsetmacro\xlambar{1-\xlam};
    \coordinate(xlag)at($\xlambar*(BL)+\xlam*(BR)$);
    \pgfmathsetmacro\ylam{(4000^(\pow)-\ymin+6)/(\ymax-\ymin+12)};
    \pgfmathsetmacro\ylambar{1-\ylam};
    \coordinate(ylag)at($\ylambar*(BL)+\ylam*(TL)$);
    \path
    ($(xlast|-ylast)$)edge[->]node[above=6pt,scale=0.9,inner sep=0pt,pos=0.05,anchor=west]{social distancing}node[below=6pt,scale=0.9,inner sep=0pt,pos=0.05,anchor=west]{lag, \math{k=10}}(xlag|-ylast)
    ($(xlock|-ylock)$)edge[->]node[above=6pt,scale=0.9,inner sep=0pt,pos=0.05,anchor=west]{lockdown}node[below=6pt,scale=0.9,inner sep=0pt,pos=0.05,anchor=west]{lag, \math{k=10}}(xlocklag|-ylock)
    ;    
    \node[scale=1,inner sep=0pt](xlabel)at($1.07*(F.south)-0.07*(F.north)$){date};
    \node[scale=1,rotate=90,inner sep=0pt](ylabel)at($1.07*(F.west)-0.07*(F.east)$){cumulative confirmed infections (\math{\times 1000})};
    \node(C)at($(F)+(-4.25,+1)$){\color{fillgray}\textsc{\begin{tabular}{c}Training\\the\\Model\end{tabular}}};
  \end{tikzpicture}
  \vspace*{-5pt}
  \caption{\small
    Training is the gray region and model predictions are the blue envelope.
    Observed
    infections fall away from predicteds, indicating that 
    social distancing is working in agreement with our lag of 10 days
    (the two ``kinks'' in the curve). The figure emphasizes 
    early data for  learning about the pandemic, as
    later data is ``contaminated'' by public health protocols whose
    effects are hard to quantify. (Note: Dates are the time-stamps on the ECDC report \citep{ECDC2020}, which captures the previous day's activity)
    \label{fig:model3}}
\end{figure}

We analyzed daily confirmed COVID-19 cases
from January 21 to March 14, the training or model calibration phase, the
gray region in Figure~\ref{fig:model3}. A more detailed plot of the model fit to
the training data is in Figure~\ref{fig:model1}.
Qualitatively we see that the model captures the data, and it does so
by setting
four parameters:
\myslimfbox[fillLightgray]{0.9\textwidth}{
\math{\arraycolsep2pt
  \begin{array}{rl}
    \beta,&\text{{\bf asymptomatic}
      infectious force governing exponential
      spread}\\
    \gamma,&\text{virulence, the fraction of mild cases that become
      serious later}\\
    k,&\text{lag time for mild infection to become serious
      (an incubation time)}\\
    M_0,& \text{Unconfirmed mild asymptomatic
      infections at time 0}
  \end{array}
}
}
Calibrating the model to the training data, gives the following
information.
\mldc{\arraycolsep6pt\renewcommand{\arraystretch}{1.1}
  \begin{array}{l|ll}
    \text{parameter}&\text{model}&\text{range}\\\hline
    \text{asymptomatic
    infectious force }\beta&\beta^*=1.30&[1.3,1.31]
\\
\text{virulence }\gamma&\gamma^*=0.14\%&[0.03\%,1.2\%]
\\
\text{lag }k&k^*=10\text{ days}&[1\text{ day},13\text{ days}]
\\
\text{initial infections }M_0&M_0^*=4&[1,12]
\\
\text{asymptomatic on 03/14 }&5.3\text{ million}&[1.3,26]\text{ million}
\end{array}
  \label{eq:model1}
}
The \emph{asymptomatic}
infectious force if left unchecked is
30\% new cases per day (doubling in about 2.6 days)
and the virulence is
\math{\gamma=0.14\%} (1 or 2 in a thousand conversions from mild infection
to serious). Not all serious cases are fatal.
The model output about 5.3 million asymptomatic cases as of 03/14 and a range
from 1 to 26 million, a surprisingly high number.
Such quantitative early intelligence
has significance for public
health protocols.

Beyond 03/14 in Figure~\ref{fig:model3} are the model
predictions (blue envelope) and the red circles are
the observed infection counts. How do we know the model predictions
are honest, in that the red circles were in no way influenced the
predictions. We are in a unique position to test the model because
it is \emph{time stamped} as  version 1 of the
preprint~\cite{malik233}.
The model has provably not changed since 03/14,
and we just added test data as it arrived.
The predictions in Figure~\ref{fig:model3}
are in no way  forward looking, data snooped or overfitted.
We observe that the model and observed counts agree,
modulo two ``kinks'' around 03/24 and 03/30, when the observed
infections start falling away from the model. To understand the cause, the
lag is important. Aggressive social distancing was implemented on about
03/13 and lockdowns around 03/21. A lag of \math{k=10} means the effects of
these protocols will become apparent around 03/23 and 03/31 respectively.

The methods are general and can be applied to different cohorts. In
Section~\ref{sec:country} we do a cross-sectional country-based study.
Our contributions are
\begin{itemize}\itemsep1pt\vspace*{-5pt}
\item A methodology for quickly and robustly machine learning simple
  epidemiological models given coarse aggregated infection counts.
\item Building a simple model with lag for learning from early
   pandemic dynamics.
\item Application of the methodology within the context of COVID-19
  to USA data. Our methods reveal significant asymptomatic
  (hidden) infection, a lag of about 10 days, and we quantitatively
  confirm that the infectious force is strong with about a 0.14\%
  transition from mild to serious infection.
\item Cross-sectional analysis of the pandemic dynamics across several
  countries.
\item To our knowledge, the \emph{only} tested predictions for COVID-19 due to
  our time-stamping of the predictions. Our results demonstrate
  the effectiveness of simple robust models for predicting pandemic
  dynamics from \emph{early} data.
\end{itemize}

\section{Model and Method}

\begin{figure}[t]
  \centering
  \begin{tikzpicture}
    \node[inner sep=0pt](F)at(0,0){\includegraphics[width=0.7\textwidth,trim={1cm 6.7cm 1.35cm 6.9cm},clip]{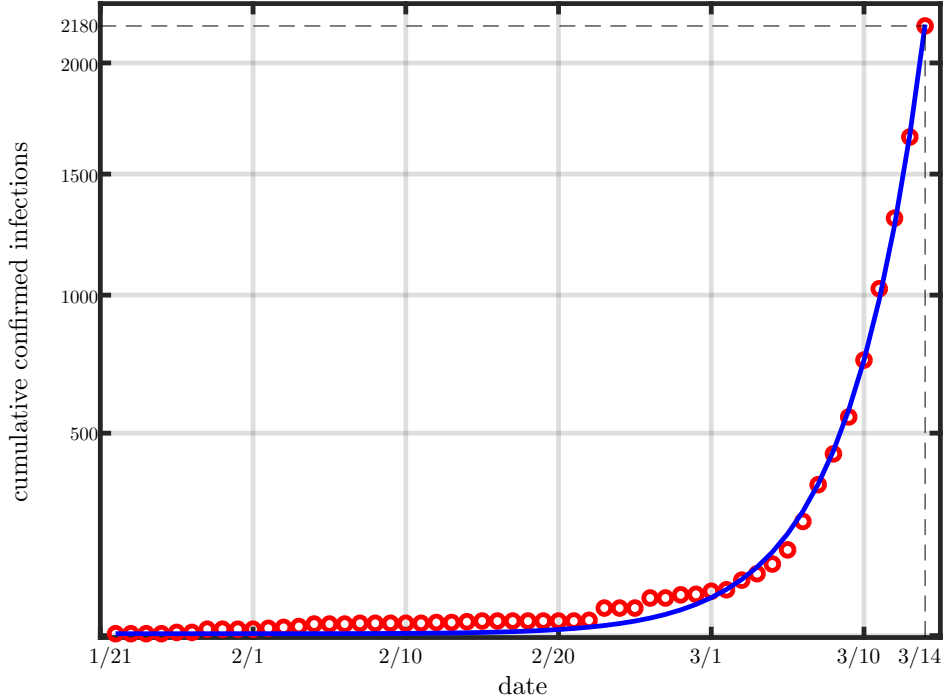}};
    \foreach\y/\x[count=\i]in{1/{1/21},10/{2/1},20/{2/10},30/{2/20},40/{3/1},50/{3/10},54/{3/14}}{
      \pgfmathsetmacro\lam{1};
      \pgfmathsetmacro\lambar{1-\lam};
      \pgfmathsetmacro\z{1-(\y-1+1.5)/(55-1+3)}
      \pgfmathsetmacro\zbar{1-\z};
      \node[scale=0.75,inner sep=0pt,anchor=north](x\i)at($\z*\lam*(F.west|-F.south)+\z*\lambar*(F.west|-F.north)+\zbar*\lam*(F.east|-F.south)+\zbar*\lambar*(F.east|-F.north)$){\x};
    }

    \foreach\y[count=\i]in{500,1000,1500,2000,2180}{
      \pgfmathsetmacro\lam{0.98};
      \pgfmathsetmacro\lambar{1-\lam};
      \pgfmathsetmacro\z{1-(\y^(0.75)+1+4)/(330.05+1+8)}
      \pgfmathsetmacro\zbar{1-\z};
      \node[scale=0.65,anchor=east](x\i)at($\z*\lam*(F.west|-F.south)+\z*\lambar*(F.east|-F.south)+\zbar*\lam*(F.west|-F.north)+\zbar*\lambar*(F.east|-F.north)$){\y};
      }
    
    \node[scale=0.9,inner sep=0pt](xlabel)at($1.06*(F.south)-0.06*(F.north)$){date};
    \node[scale=0.9,rotate=90,inner sep=0pt](ylabel)at($1.08*(F.west)-0.08*(F.east)$){cumulative confirmed infections};    
  \end{tikzpicture}
  \vspace*{-5pt}
  \caption{\small
    Model calibration to the early dynamics (first 54 infection counts)
    of (USA) COVID-19.
    Dates are the time-stamps on the ECDC report, which captures the previous day's activity (e.g. the time stamp 1/21 is the infection on 1/20).\label{fig:model1}}
\end{figure}

Our model is simple and robust.
The majority of disease models
aggregate individuals according to disease status,
such as SI, SIR, SIS, \cite{Kermack1927,Bailey1957,Anderson1992}.
We use a similar model by considering a mild infected state which can
transition to a serious state.
Early data allows us to make
simplifying assumptions. In the early phase, when public health
protocols have not kicked in, a confirmed infection is self-reported. That is,
\emph{you} decide to
get tested. Why? Because the condition became serious. This is
important. A confirmed case is a transition from
mild infection to serious. This is not true later when, for example,
public health
protocols may institute randomized testing. At
time \math{t} let there be \math{C(t)} confirmed cases and correspondingly
\math{M(t)} mild unreported asymptomatic infections. The new confirmed cases
at time \math{t} correspond to mild infections at some earlier time
\math{t-k} which have now transitioned to serious and hence
got self-reported.
Let a fraction \math{\gamma} of those mild
cases transitioned to serious,
\mandc{C(t)=C(t-1)+\gamma M(t-k).}
Another
advantage of early dynamics is that we may approximate the growth from each
infection as independent and exponential, according to the infectious force
of the disease. So,
\mandc{M(t)=\beta M(t-1)-\gamma M(t-k)-(1-\gamma)M(t-k-r)+\alpha C(t-1).}
Here, the second term is the loss of mild cases that transitioned to
serious, the third term is the remaining 
cases that don't transition to serious
recovering at some later time \math{r}
  and the fourth term accounts for
new infections from
confirmed cases. We will further simplify and assume that confirmed cases
are fully quarantined and so \math{\alpha=0} and recovery to a non-infectious
state occurs
late enough to not factor into early dynamics. Our simplified model is:
\mldc{
  \begin{array}{rcl@{\hspace*{0.5in}}l}
  S(t)&=&S(t-1)+\gamma M(t-k)& S(t)=S(1)\text{ for } 1\le t<k\\
  M(t)&=&\beta M(t-1)-\gamma M(t-k)& M(1)=M_0.
  \end{array}
}
We set \math{t=1} at the first confirmed infection
(Jan 21 in USA). Given \math{k,M_0}, we get an approximate fit to the data
by using a perturbation analysis to solve for
\math{\gamma,\beta} that fit two points
\math{S(\tau)} and \math{S(T)}:
\eqarc{
  \gamma&\approx&
    \frac{(\phi-1)((\phi^r-1)\Delta_T+(\phi^s-1)\Delta_\tau)}{
      (\phi^r-1)^2+(\phi^s-1)^2}
    \\
    \beta&\approx&\phi\left(1+\frac{\gamma}{\phi^k+(k-1)\gamma}\right),
}
where,
\mld{
    \phi\approx
    \kappa^{1/(r-s)}\left(1-\frac{(\rho/\kappa)(r-s)}{(r-s)\kappa^{s/(r-s)}-(\rho/\kappa)s}
    \right)^{1/(r-s)}
  }
and \math{r=T-k}, \math{s=\tau-k},
  \math{\kappa=(S(T)-S(1))/(S(\tau)-S(1))} and
  \math{\rho=\kappa-1} (for details see the appendix).
  From this solution as a starting point, we can further optimize
  \math{\gamma,\beta} using a gradient descent which minimizes 
  an error-measure that captures how well 
the parameters \math{\beta,\gamma,k,M_0} 
reproduce the
observed dynamics in Figure~\ref{fig:model1}, see
for example \cite{malik173}.
We used a combination of
root-mean-squared-error  and root-mean-squared-percentage-error
between observed
dynamics and model predictions.
By optimizing over
\math{k,M_0}, we obtain an optimal fit to the training data
(Figure~\ref{fig:model1}) using model parameters:
\mldc{
  \beta^*=1.30\qquad\qquad
  \gamma^*=0.0014 \qquad\qquad
  k^*=10\text{ days}      \qquad\qquad
  M_0^*=4\label{eq:param-pot}
}
The asymptomatic infectious force, \math{\beta}, is very high, and corresponds
to a doubling time of 2.6 days. The
virulence at 0.14\% seems comparable to a standard flu, though the virus
may be affecting certain demographics much more severely than a flu.
The incubation period of 10 days seems in line with physician
observations. The data analysis predicts that when
the first confirmed case appeared, there were 4 other infections in the USA.
The parameters \math{\beta^*, \gamma^*} and \math{M_0^*} are new knowledge,
gained with relative ease by calibrating a simple robust
model to the early dynamics. But, these optimal parameters are not the whole
story, especially when it comes to prediction.

\begin{figure}[t]
  {\tabcolsep0pt
  \begin{tabular}{c@{\hspace*{0.025\textwidth}}c}
  \begin{tikzpicture}
    \node[inner sep=0pt](F)at(0,0){\includegraphics*[width=0.45\textwidth,trim={4.85cm 8cm 4.35cm 7.5cm},clip]{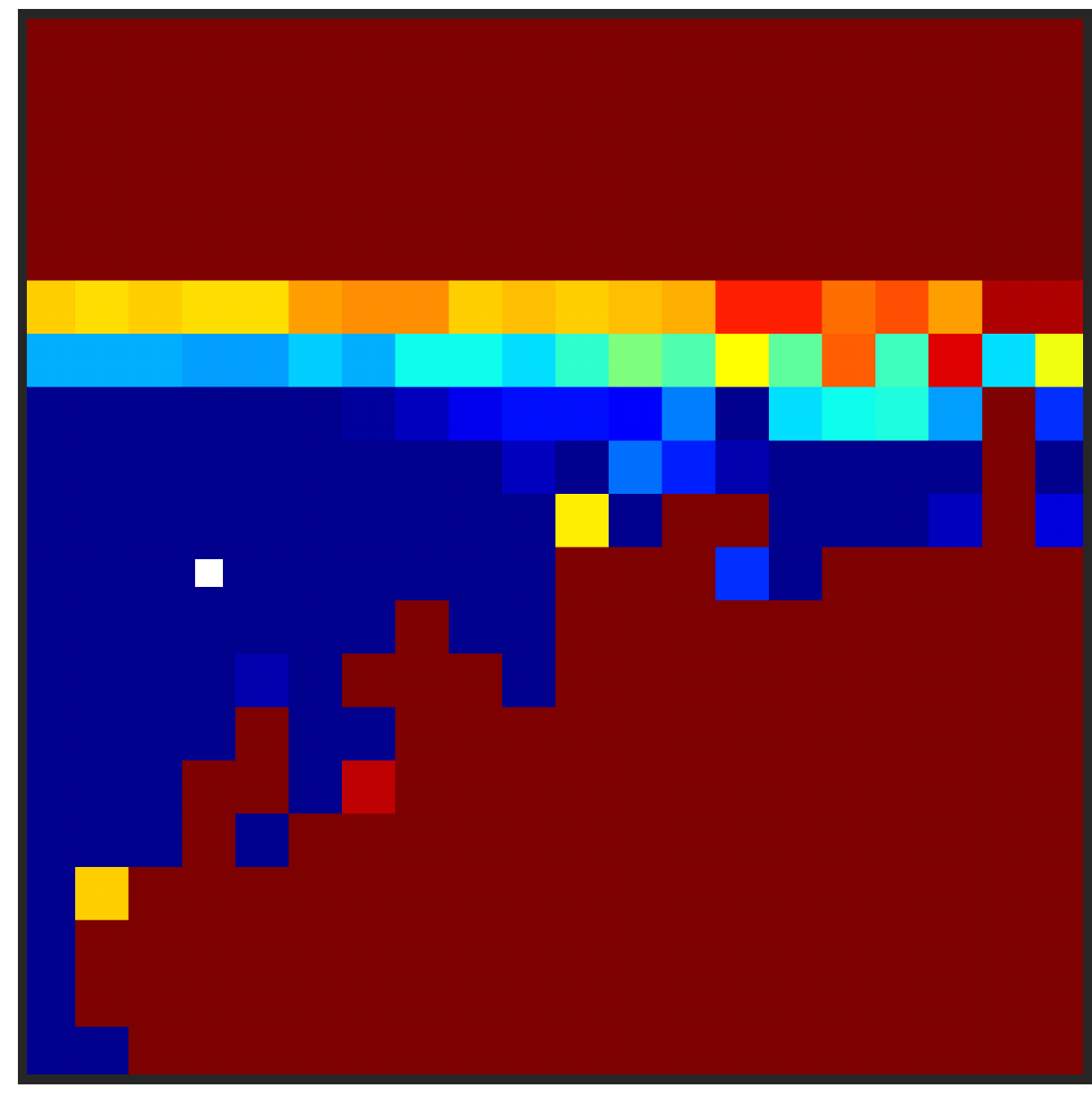}};
    \foreach\x/\z[count=\i]in{{1}/0.98,{20}/0.02}{
      \pgfmathsetmacro\lam{1};
      \pgfmathsetmacro\lambar{1-\lam};
      \pgfmathsetmacro\zbar{1-\z};
      \node[scale=0.8,inner sep=0pt,anchor=north](x\i)at($\z*\lam*(F.west|-F.south)+\z*\lambar*(F.west|-F.north)+\zbar*\lam*(F.east|-F.south)+\zbar*\lambar*(F.east|-F.north)$){\x};
    }
    \foreach\y/\z[count=\i]in{{1}/0.97,{20}/0.03}{
      \pgfmathsetmacro\lam{0.98};
      \pgfmathsetmacro\lambar{1-\lam};
      \pgfmathsetmacro\zbar{1-\z};
      \node[scale=0.8,anchor=east](x\i)at($\z*\lam*(F.west|-F.south)+\z*\lambar*(F.east|-F.south)+\zbar*\lam*(F.west|-F.north)+\zbar*\lambar*(F.east|-F.north)$){\y};
    }
    \node[scale=0.9,inner sep=0pt](xlabel)at($1.04*(F.south)-0.04*(F.north)$){\math{M_0}, initial mild infections};
    \node[scale=0.9,rotate=90,inner sep=0pt](ylabel)at($1.03*(F.west)-0.03*(F.east)$){\math{k}, incubation period (days)};
  \end{tikzpicture}
  &
   \begin{tikzpicture}
    \node[inner sep=0pt](F)at(0,0){\includegraphics*[width=0.45\textwidth,trim={4.85cm 8cm 4.35cm 7.5cm},clip]{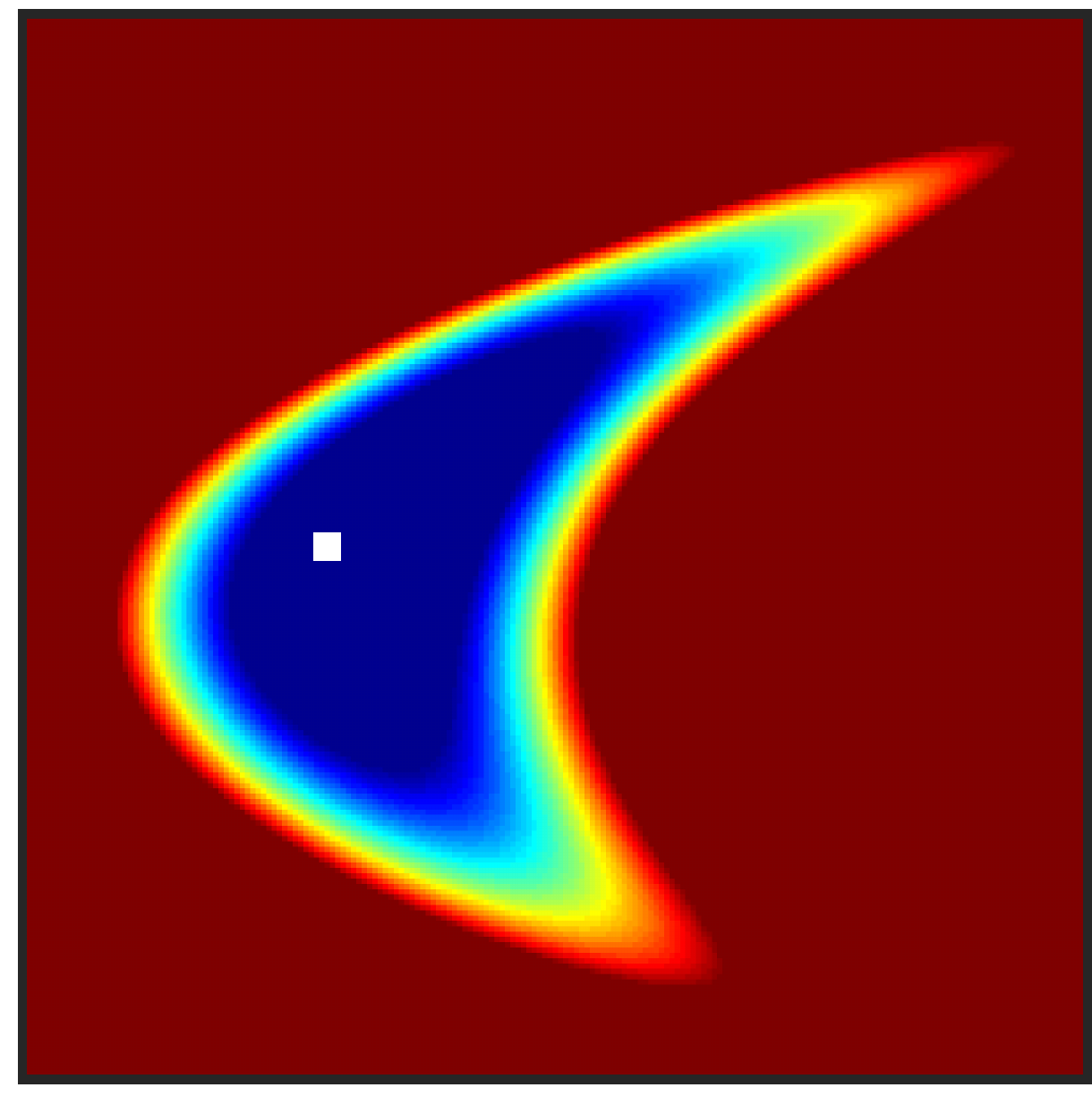}};
    \foreach\x/\z[count=\i]in{{-0.46}/0.98,{1.14}/0.02}{
      \pgfmathsetmacro\lam{1};
      \pgfmathsetmacro\lambar{1-\lam};
      \pgfmathsetmacro\zbar{1-\z};
      \node[scale=0.8,inner sep=0pt,anchor=north](x\i)at($\z*\lam*(F.west|-F.south)+\z*\lambar*(F.west|-F.north)+\zbar*\lam*(F.east|-F.south)+\zbar*\lambar*(F.east|-F.north)$){\x};
    }
    \foreach\y/\z[count=\i]in{{-18}/0.97,{18}/0.03}{
      \pgfmathsetmacro\lam{0.98};
      \pgfmathsetmacro\lambar{1-\lam};
      \pgfmathsetmacro\zbar{1-\z};
      \node[scale=0.8,anchor=east](x\i)at($\z*\lam*(F.west|-F.south)+\z*\lambar*(F.east|-F.south)+\zbar*\lam*(F.west|-F.north)+\zbar*\lambar*(F.east|-F.north)$){\y};
    }
    \node[scale=0.9,inner sep=0pt](xlabel)at($1.04*(F.south)-0.04*(F.north)$){\math{1400\beta+35\gamma}};
    \node[scale=0.9,rotate=90,inner sep=0pt](ylabel)at($1.03*(F.west)-0.03*(F.east)$){\math{1400\beta-35\gamma}};
   \end{tikzpicture}
   \\
   (a)&(b)
  \end{tabular}
  }
  \vspace*{-8pt}
  \caption{\small Uncertainty in model estimation. (a) Optimal fit-error over
    for different choices of \math{M_0} and \math{k}.
    Blue is better fit, red is worse. The deep-blue
    region corresponds to comparable models, within 0.5\% of the
    optimal fit.
    The white square is the model chosen, \math{(k^*,M_0^*)=(10,4)}
    which has optimal fit and also is
    ``robust'' being in the middle of the deep-blue regions.
    (b) Model fit for the chosen \math{k^*} and \math{M_0^*}.
    Again, the deep-blue is an equivalence
    class of optimal models. The ``robust'' model
    is the
    white square in the middle of the deep-blue region,
    \math{(\beta^*,\gamma^*)=(1.30,0.0014)}.
    The deep-blue regions represent uncertainty.\label{fig:model2}}
\end{figure}
The exhaustive search over \math{k,M_0},
fixing \math{\beta} and \math{\gamma} to the optimal for
that specific \math{k,M_0}, produces several equivalent models
We show the quality of fit
for various \math{(k,M_0)} in Figure~\ref{fig:model2}(a).
The deep-blue region contains essentially equivalent models within
0.5\% of the optimal fit, our (user defined) error tolerance.
The deep-blue region shows the extent to which the
model is ill-identified by the data. Indeed, all these deep-blue
models equally
fit the data which results in a range of predictions. For robustness, we
pick the white square in the middle of the deep-blue region, but note here that it
is only one of the models which are consistent
with the data. In making predictions, we
should consider all equivalent models to get a range for the predictions that
are all equally consistent with the data.
Similarly, in Figure~\ref{fig:model2}(b), we fix
\math{k^*} and \math{M_0^*} to their optimal robust values
and show the uncertainty with respect to
\math{\beta} and \math{\gamma} (the deep-blue region). Again, we pick the white square
in the middle of the deep-blue region of equivalent
models with respect to the data. Hence we arrive at our optimal
parameters in Equation \r{eq:param-pot}. 
By considering all models which are equally consistent with the data, we
get the estimates toghether with the ranges in Equation~\ref{eq:model1}.
We emphasize that these error-ranges we report have nothing to
do with the data, and are simply due to the
ill-posedness of the inverse problem to inifer the model from
finite data. Several models essentially fit the data equivalently.
We do not include in our range the possible
measurement errors in the data, although the two are related
through the error tolerence used in defining ``equivalent'' models.
More noise in the data would result in more models being treated as equivalent.

\section{Results}
As already mentioned, to get honest estimates, we must
consider all models which are equally consistent with the data
(deep-blue regions in Figure~\ref{fig:model2}).

\subsection{COVID-19 in USA}
The model in Equation~\r{eq:model1} gives the
prediction of new infections in the table below. The cumulative
predicted infections is the
data plotted in Figure~\ref{fig:model3}.
\mandc{\scalebox{0.8}{\text{\tabcolsep10pt\renewcommand{\arraystretch}{1.1}
  \begin{tabular}{c|cl|c}
    &\multicolumn{3}{c}{New Infections}\\
    Date&Model Prediction&\multicolumn{1}{l}{Range}&Observed\\\hline
    March 15, 2020&665&[657,2861]&\blue{\bf 777}\\
    March 16, 2020&866&[856,3732]&\red{\bf 823}\\
    March 17, 2020&1130&[1114,4867]&\red{\bf 887}\\
    March 18, 2020&1475&[1450,6348]&\blue{\bf 1766}\\
    March 19, 2020&1924&[1888,8280]&\blue{\bf 2988}\\
    March 20, 2020&2510&[2459,10800]&\blue{\bf 4835}\\
    March 21, 2020&3275&[3201,14088]&\blue{\bf 5374}\\
    March 22, 2020&4272&[4168,18375]&\blue{\bf 7123}\\
    March 23, 2020&5574&[5427,23968]&\blue{\bf 8459}\\
    March 24, 2020&7271&[7067,31263]&\blue{\bf 11236}\\
    March 25, 2020&9487&[9201,40789]&\red{\bf 8789}\\
    March 26, 2020&12376&[11980,53216]&\blue{\bf 13693}\\
    March 27, 2020&16146&[15198,69430]&\blue{\bf 16797}\\
    March 28, 2020&21065&[20309,90585]&\red{\bf 18695}\\
    March 29, 2020&27482&[26443,118180]&\red{\bf 19979}\\
    March 30, 2020&35854&[34430,154190]&\red{\bf 18360}\\
    March 31, 2020&46777&[44829,201170]&\red{\bf 21595}\\
    April 1, 2020&61026&[58369,262470]&\red{\bf 24998}\\
    April 2, 2020&79617&[75999,343440]&\red{\bf 27103}\\
\end{tabular}}}}
The predictions use the model in \r{eq:model1}
and the ranges are obtained by using
the space of models equally consistent with the data. March 15 and 16 data
  arrived at the time of writing and March 17 onward arrived after the time of
  writing time-stamped version 1 \citep{malik233}.
  Blue means in the range and red means outside the range.

\begin{table}[!t]  
\centering
\scalebox{0.9}{  \tabcolsep5pt
  \begin{tabular}{l||lc|lc|lc|lc}
    Country&\math{\beta}&\math{[{\min},{\max}]}&\math{\gamma}&\math{[{\min},{\max}]}&\math{k}&\math{[{\min},{\max}]}&\math{M_0}&\math{[{\min},{\max}]}\\\hline
Australia&1.143&[1.137,1.18]&0.0047&[0.0024,0.046]&2&[1,3]&10&[1,18]\\
Austria&1.411&[1.389,2.261]&0.045&[0.022,0.89]&1&[1,1]&20&[1,40]\\
Canada&1.182&[1.176,1.191]&0.0014&[0.00025,0.0099]&3&[1,3]&10&[1,40]\\
China&1.312&[1.297,1.332]&5.18&[3.75,8.26]&15&[14,16]&6&[5,9]\\
France&1.297&[1.293,1.301]&0.05&[0.0128,0.449]&19&[18,20]&7&[1,36]\\
Germany&1.289&[1.287,1.295]&0.011&[0.0039,0.044]&8&[1,8]&4&[1,10]\\
Iran&1.596&[1.563,1.709]&0.78&[0.71,0.79]&4&[3,4]&30&[29,40]\\
Ireland&1.401&[1.372,1.79]&0.083&[0.042,0.62]&2&[2,2]&20&[2,40]\\
Italy&1.242&[1.237,1.255]&3.88&[3.11,5.37]&19&[18,19]&6&[4,8]\\
Netherlands&1.387&[1.372,2.28]&0.34&[0.29,1.314]&4&[2,4]&35&[2,40]\\
Norway&1.461&[1.379,2.935]&0.164&[0.082,1.64]&1&[1,3]&20&[1,40]\\
Poland&1.476&[1.422,3.542]&0.109&[0.054,2.17]&1&[1,1]&20&[1,40]\\
Portugal&1.4&[1.356,3.077]&0.0886&[0.044,1.77]&1&[1,1]&20&[1,40]\\
South Africa&1.563&[1.513,2.65]&0.057&[0.031,1.15]&1&[1,2]&20&[1,37]\\
South Korea&1.238&[1.231,1.289]&0.98&[0.76,4.65]&20&[14,20]&7&[1,9]\\
Spain&1.411&[1.405,1.418]&0.127&[0.016,0.63]&20&[20,20]&5&[1,40]\\
Sweden&1.356&[1.343,1.372]&0.064&[0.0068,0.27]&19&[19,20]&3&[1,40]\\
Switzerland&1.527&[1.448,2.121]&0.4&[0.267,1.078]&3&[2,3]&12&[2,21]\\
UK&1.252&[1.248,1.257]&0.68&[0.412,0.878]&19&[17,20]&1&[1,1]\\
USA&1.306&[1.303,1.309]&0.0014&[0.0006,0.009]&10&[3,12]&4&[1,7]
\end{tabular}}
\medskip
\caption{Fit parameters for 20 countries.\label{tab:intro:country-fits}}
\end{table}
\subsection{Cross-Sectional Study By Countries}
\label{sec:country}

In the supplementary material we give details of our
cross-sectional study across
countries. The different countries have different
cultures, social networks, mobility, demographics, as well as different
times at which the first infection was reported (the ``delay'').
We calibrated independent models for each country and the
resulting model parameters are 
in Table~\ref{tab:intro:country-fits}.

We primarily focused on 
the infectious force \math{\beta}, which has significant variability,
and we studied how \math{\beta} statistically depends on a number of
country-specific parameters factors. In the supplementary material, we
give details of the study and the quantitative results.  Qualitatively, we
find:
\begin{itemize}\itemsep1pt\vspace*{-5pt}
\item
  A larger delay in the virus reaching a country indicates 
  a larger \math{\beta}. The more that has been
  witnessed, the faster the spread. That seems unusual, but
  is strongly indicated. We do not have a good explanation for
  this effect. It could be an artefact of testing procedures not being
  streamlined, so early adopters of the pandamic presenting as serious were
  not detected.
  
\item Population density at the infection site has a strong
  positive effect but the country's population density does not.
\item
  There is faster spread in countries with more people under the
  poverty level defined as the percentage of people
  living on less than \$5.50 per day.
\item
  Median age has a strong effect. Spread is faster in younger populations.
  The youth are more mobile and perhaps also more carefree.
\item Wealth and per-capita income have a slight negative effect.
  Spread is slower in richer countries, perhaps due to
  risk-aversion, higher levels of education and less
  reliance on public transportation. Whatever the cause, it does have an
  impact, but relatively smaller than the other effects.
  \vspace*{-5pt}
\end{itemize}


\section{Conclusion}

Early dynamics allows us to learn useful properties of the pandemic.
Later dynamics may be contaminated by human intervention,
which renders the data
less useful without a more general model.
We learned a lot from the early dynamics of COVID-19.
It's infectious
force, virulence, incubation period, unseen infections and predictions of
new confirmed cases. All this, albeit within error tolerances, from a
simple model and little data.
Asymptomatic
infection is strong, around 30\%,
converting to serious at a rate at most 1.2\%.
There is significant uncertainty in the lag, from 1 up to
13 days, and we estimate 5.3 million asymptomatic infections as of 03/14, the
range being from 1 to 26 million.
Such information is useful for
planning and health-system preparedness.
Are our parameters
correct?
We were in a unique position to \emph{test} our predictions because our
model was \emph{time-stamped} as  version 1 of the
preprint~\cite{malik233}. 

A side benefit of the model predictions is as a benchmark 
against which to evaluate public health interventions. If moving forward,
observed new infections are low compared to the data in, it means the
interventions are working by most likely
reducing \math{\beta}. Starting on about March 25, the observed infections
starts falling off and we observe a flattening by March 28. 
The US instuted broad and aggressive social distancing protocols starting on
or before March 13 and even
stronger lockdown around March 21, which is consistent
with the data and the model's lag of \math{k=10}.
Without such quantitative targets to compare with, it would
be hard to evaluate intervention protocols.

Our approach is simple
and works with coarse, aggregated  data.
But, there are 
limitations.
\begin{itemize}\itemsep-2pt\vspace*{-5pt}
\item
  The independent evolution of infection sites only applies
to early dynamics.  Hence,
when the model infections increase beyond some point, and
the pandemic begins to saturate the population,
a more sophisticated network model that
captures the dependence between infection sites
would be
needed~\cite{Balcan2009,Hill2010,Salathe2010,Keeling2005,Chinazzi2020}.

\item
  While we did present an optimal model, it should be taken with
a grain of salt because many models are nearly equivalent, resulting in 
prediction uncertainty.

\item
  The model and the
interpretation of its parameters will change once public health
protocols kick in. The model may have to be re-calibrated
(for example if \math{\beta} decreases) and the parameters may have to
be reinterpreted (for example \math{\gamma} is a virulence only in the
self-reporting setting, not for random testing).
It is also possible to build a more general model with an early phase
\math{\beta_1} and a latter phase \math{\beta_2}
(after social distancing).
But, beware, for a more general sophisticated model looks
good \emph{a priori} until it comes time to calibrate it to data, at which
point it becomes unidentifiable.

\item The model was learned
on USA data. The learned model parameters may not be
appropriate for another society. The virulence could be globally
invariant, but it could also depend on genetic and demographic factors
like age, as well as what ``serious'' means for the culture - that is when
do you get yourself checked. In a high-strung society, you expect
high virulence-parameter
since the threshold for serious is lower. One certainly expects 
the infectious force to depend on the underlying social network and
interaction patterns between people, which can vary drastically from one
society to another and depending on interventions.
Hence, one should
calibrate the model to country specific data to gain more useful insights.
\vspace*{-5pt}
\end{itemize}
\emph{The Lag, \math{k}, and Public Policy.}
The lag \math{k} is
important for public policy due to how public policy can be
driven by human psychology.
The Human tendency is to associate
any improvement in outcome to recent actions. However, if there is
a lag, one might prematurely reward those recent actions
instead of the earlier actions whose effects are actually being seen.
Such lags are present in traditional machine learning, for example
the delayed reward in reinforcement learning settings. Credit assignment
to prior actions in the face of delayed reward is a notoriously difficult
task, and this remains so with humans in the loop. Knowledge of the lag
helps to assign credit appropriately to prior actions, and the
public health setting is no exception.


\paragraph{Acknowledgments.}
We thank Abhijit Patel, Sibel Adali and Zainab Magdon-Ismail for
provoking discussions on this topic.

{\small
\bibliographystyle{natbib}
\bibliography{mypapers,masterbib,covid} 
}

\newpage
\appendix

\section{Fitting The Model}

Recall the model,\mldc{
  \begin{array}{rcl@{\hspace*{0.5in}}l}
  S(t)&=&S(t-1)+\gamma M(t-k)& S(t)=S(1) \text{ for } 1\le t<k\\
  M(t)&=&\beta M(t-1)-\gamma M(t-k)& M(1)=M_0, M(t)=0 \text{ for }t\le0.
  \end{array}
}
For fixed \math{k,M_0}, we must perform a gradient descent to
optimize \math{\beta,\gamma}. Unfortunately, the dependence on \math{\beta} is
exponential and hence very sensitive. So
if the starting point is not chosen carefully, the optimization gets stuck
in a very flat region, and many millions of iterations are
needed to converge. Hence it is prudent to choose the starting
conditions carefully. To do so, we need to analyze the recursion.
First, we observe that the recursion for
\math{M(t)} is a standalone \math{k}-th order recurrence.
For \math{1\le t\le k}, \math{M(t)=M_0\beta^{t-1}}, 
hence, we can
guess a solution \math{M(t)=M_0\beta^{k-1}\phi^{t-k}}, for \math{t> k},
which requires
\mandc{
  \phi^k-\beta\phi^{k-1}+\gamma=0.}
We do a perturbation analysis in \math{\gamma\rightarrow0}.
At \math{\gamma=0}, \math{\phi=\beta}, so we set
\math{\phi=\beta+\epsilon}, to get
\mandc{
  (\beta+\epsilon)^k-\beta(\beta+\epsilon)^{k-1}+\gamma=0,}
which to first order in \math{\epsilon} is solved by
\math{\epsilon\approx -\gamma/\beta^{k-1}} and so
  \mandc{
    M(t)\approx
    \begin{cases}
    M_0\beta^{t-1}&1\le t\le k,\\
    M_0\beta^{k-1}\phi^{t-k}&t>k,
    \end{cases}
  }
  where \math{\phi=\beta(1-\gamma/\beta^k)}.
  Given this approximation, we can solve for
  \math{S(t)},
  \mandc{
    S(t)=
  \begin{cases}
    \displaystyle S(1)&1\le t\le k;\\
    \displaystyle S(1)+\frac{\gamma M_0(\beta^{t-k}-1)}{\beta-1}&k<t\le 2k\\
    \displaystyle S(1)+\frac{\gamma M_0(\beta^{k}-1)}{\beta-1}+\frac{\gamma M_0\beta^{k-1}\phi(\phi^{t-2k}-1)}{\phi-1}&2k<t.
   \end{cases}
}%
Since \math{\phi=\beta+O(\gamma)}, for \math{t>2k}, we can
approximate \math{S(t)} as,
  \mandc{
    S(t)\approx S(1)+\frac{\gamma M_0(\phi^{t-k}-1)}{\phi-1}.
  }%
We can independently control two parameters
  \math{\phi} and \math{\gamma}. We use this to match the observed
  \math{S(t)} at two time points. Since the growth is
  exponential, we match the end time, \math{S(T)} and
  some time \math{\tau} in the middle, for example
  \math{\tau=\ceil{3T/4}}. Let
  \math{\Delta_T=(S(T)-S(1))/M_0} and \math{\Delta_\tau=(S(\tau)-S(1))/M_0}.
  Then,
  \eqarc{
      \Delta_T&=&\frac{\gamma (\phi^{T-k}-1)}{\phi-1};\\
      \Delta_\tau&=&\frac{\gamma (\phi^{\tau-k}-1)}{\phi-1}.
      \label{eq:app:1}
  }
  Dividing gives
  \math{\Delta_T/\Delta_\tau=(\phi^{T-k}-1)/(\phi^{\tau-k}-1)\approx
    \phi^{T-\tau}}, because \math{\phi>1}.
  Let us consider the equation
  \math{\kappa=(\phi^r-1)/(\phi^s-1)}, which gives
  \math{\phi^r-\kappa\phi^s+\kappa-1=0}, or more generally
  \math{\phi^r-\kappa\phi^s+\rho=0}, where
  \math{r>s>1} and \math{\kappa>\rho\gg 1}.
  This means \math{\phi>1}. When \math{\rho=0}, we have
  \math{\phi^{r-s}=\kappa}, so we do a perturbation analysis
  with \math{\phi^{r-s}=\kappa+\epsilon}, and our perturbation parameter
  is \math{\epsilon}. Then, \math{\phi^r=(\kappa+\epsilon)\phi^s} and plugging
  into the equation gives
  \mandc{
    \epsilon=-\frac{\rho}{\phi^{s}}=-\frac{\rho}{(\kappa+\epsilon)^{s/(r-s)}}
    \approx
    -\frac{\rho}{\kappa^{s/(r-s)}}\left(1-\frac{s}{r-s}\frac{\epsilon}{\kappa}\right).
  }
  Solving for \math{\epsilon} gives
  \math{\epsilon\approx-\rho(r-s)/((r-s)\kappa^{s/(r-s)}+s)}, which gives
  \mldc{
    \phi\approx
    \kappa^{1/(r-s)}\left(1-\frac{(\rho/\kappa)(r-s)}{(r-s)\kappa^{s/(r-s)}-(\rho/\kappa)s}
    \right)^{1/(r-s)}.
  }
  For our setting, \math{r=T-k}, \math{s=\tau-k},
  \math{\kappa=\Delta_T/\Delta_\tau} and
  \math{\rho=\kappa-1}.
  Finally, since \math{\phi} is approximate, we may not be able
  to satisfy both equations in \r{eq:app:1}, hence we
  can instead minimize the
  mean squared error, which gives
  \mldc{
    \gamma=
    \frac{(\phi-1)((\phi^r-1)\Delta_T+(\phi^s-1)\Delta_\tau)}{
      (\phi^r-1)^2+(\phi^s-1)^2}.\label{eq:app:gam}
  }
  We now need to get \math{\beta} which satisfies
  \math{\phi=\beta(1-\gamma/\beta^k)}. Again, we do a perturbation analysis,
  omitting the details, to obtain
  \mldc{
    \beta\approx\phi\left(1+\frac{\gamma}{\phi^k+(k-1)\gamma}\right).\label{eq:app:bet}
  }
  If one wishes, a fixed point iteration starting at the above will quickly
  approach a solution to \math{\phi=\beta(1-\gamma/\beta^k)}.

  \begin{figure}[t]
    \centering
    \includegraphics[width=0.6\textwidth,trim={0cm 0cm 8.1cm 16.25cm},clip]{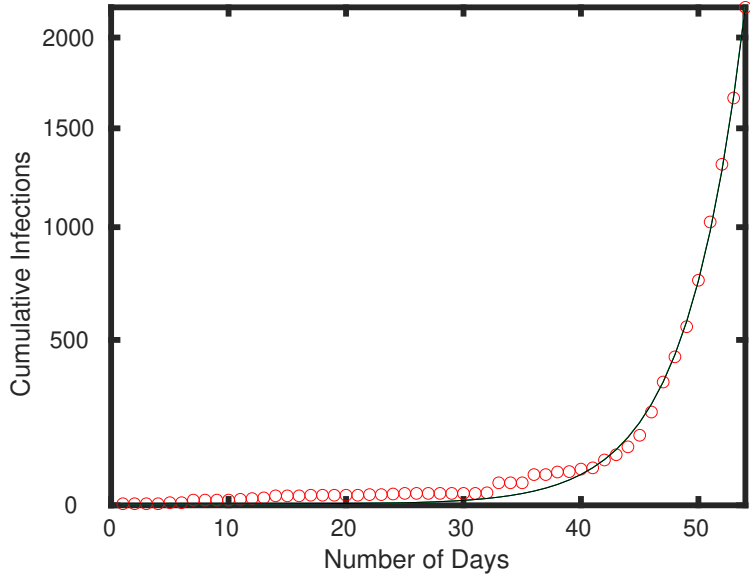}
    \caption{Approximate fit to USA data is essentially on top of optimal fit.
    \label{fig:app:approx}}
\end{figure}

  We show the approximate fit on the US data
  (Figure~\ref{fig:app:approx}).
  We show the optimal fit, the initial fit using the parameters constructed
  from 
  \r{eq:app:bet} and \r{eq:app:gam}. The parameters and fit error are
  \mandc{
    \begin{array}{r|ccc}
      &\beta&\gamma&\text{fit error}\\\hline
      \text{optimal}&1.306&0.0013282&1.6521\\
      \text{Equations \r{eq:app:bet} and \r{eq:app:gam}}
      &1.3055 & 0.0013423  &  1.6619
    \end{array}
  }
  The approximate fit works pretty well, and is certainly good enough
  to initialize an optimization. Note that to get an even better starting
  point for the gradient descent optimization, assuming
  \math{T,\tau\ge 2k}, one could simultaneously solve the three
  equations
  \eqar{
  0&=&\phi^k-\beta\phi^{k-1}+\gamma\\
  \Delta_T&=&
  \frac{\gamma M_0(\beta^{k}-1)}{\beta-1}+\frac{\gamma M_0\beta^{k-1}\phi(\phi^{T-2k}-1)}{\phi-1}\\
  \Delta_\tau&=&
  \frac{\gamma M_0(\beta^{k}-1)}{\beta-1}+\frac{\gamma M_0\beta^{k-1}\phi(\phi^{\tau-2k}-1)}{\phi-1}.
  }

\section{Cross-Sectional Country Study}

We perform our analysis on early dynamics data available from the ECDC giving
infection numbers starting from December 31, 2019 \cite{ECDC2020}.
We use data 
for 20 countries selected qualitatively because they appear to have
reasonably efficient testing procedures for self-reported cases.
We iniclude China data for completeness, even though China dynamics since
December 31 are not early dynamics.
We show these countries below, together with some demographic
data which might determine spread dynamics:
Start City; Start and End dates; Delay to first infection in days;
Population Density; Median Age \citep{Wiki};
Wealth as defined by adjusted net national income per capita~\citep{WB};
Average Income~\citep{WDinfo}; Poverty Level~\citep{WikiPoverty} defined by
living on less than \$5.50 per day; Population Density around
the first infection site for the country.

\begin{table}[!ht]
  \centering
\scalebox{0.85}{\tabcolsep6pt
\begin{tabular}{llllllllllll}
Country&Code&Start City&Start&End&Delay&Den&Age&Wealth&Income&Pov.&Den-Init\\\hline
Australia&AU&Melb./Syd.&1/25&3/14&26&3&38.7&41489&53230&1.2&425\\
Austria&AT&Innsbr./Vien.&2/26&3/14&58&106&44&38748&49310&0.9&3000\\
Canada&CA&Toronto&1/26&3/14&27&4&42.2&36802&44940&1&4150\\
China&CN&Wuhan&12/31&2/5&1&145&37.4&6568&9460&27.2&1200\\
France&FR&Bordeaux&1/25&3/14&26&123&41.4&32672&41080&0.2&5000\\
Germany&DE&Coppingen&1/28&3/14&29&233&47.1&37791&47090&0.2&972\\
Iran&IR&Qom/Tehran&2/20&3/14&52&54&30.3&4238&5470&11.6&15500\\
Ireland&IE&Dublin&3/3&3/14&64&70&36.8&37988&61390&0.7&4588\\
Italy&IT&Rome&1/31&3/14&32&200&45.5&26537&33730&3.5&2232\\
Netherl.&NL&Tilburg&2/28&3/14&60&420&42.6&40545&51260&0.5&1852\\
Norway&NO&Tr. Finn/Oslo&2/27&3/14&59&17&39.2&61865&80610&0.2&1400\\
Poland&PL&Zielona Gora&3/6&3/14&67&123&40.7&11650&14100&2.1&504\\
Portugal&PT&Porto&3/3&3/14&64&112&42.2&17188&21990&3&6900\\
S. Africa&ZA&Gaut./Dur.&3/8&3/14&69&48&27.1&4942&5750&57.1&4000\\
S. Korea&KR&Daegu&1/20&3/4&21&517&41.8&24028&30600&1.2&2818\\
Spain&ES&La Gom/Ten.&2/1&3/14&33&93&42.7&23216&29340&2.9&250\\
Sweden&SE&Jonkoping&2/1&3/14&33&23&41.2&45149&55490&1&2100\\
Switzer.&CH&Tin./Bas./Zur.&2/26&3/14&58&208&42.4&64307&84410&0&6000\\
UK&UK&Newcas.Tyme&1/31&3/14&32&274&40.5&34171&41770&0.7&233\\
USA&US&Seat./Snohom.&1/21&3/14&22&34&38.1&51485&63080&2&3430
\end{tabular}}
\medskip
\caption{Comparison of countries used in the study.\label{tab:app:country-comp}}
\end{table}

From the public health perspective, perhaps the most important
parameter is \math{\beta}, since actions can be taken to
mitigate the spread by reducing \math{\beta}, whereas
\math{\gamma, k} and \math{M_0} are somewhat givens for the country.
We show the fits in Table~\ref{tab:app:country-fits}. As you can see, there
is much variability in \math{\beta}.

\begin{table}[t]  
\centering
\scalebox{0.85}{\tabcolsep8pt
  \begin{tabular}{l||cl|cl|cl|cl}
    Country&\math{\beta}&\math{[{\min},{\max}]}&\math{\gamma}&\math{[{\min},{\max}]}&\math{k}&\math{[{\min},{\max}]}&\math{M_0}&\math{[{\min},{\max}]}\\\hline
Australia&1.143&[1.137,1.18]&0.0047&[0.0024,0.046]&2&[1,3]&10&[1,18]\\
Austria&1.411&[1.389,2.261]&0.045&[0.022,0.89]&1&[1,1]&20&[1,40]\\
Canada&1.182&[1.176,1.191]&0.0014&[0.00025,0.0099]&3&[1,3]&10&[1,40]\\
China&1.312&[1.297,1.332]&5.18&[3.75,8.26]&15&[14,16]&6&[5,9]\\
France&1.297&[1.293,1.301]&0.05&[0.0128,0.449]&19&[18,20]&7&[1,36]\\
Germany&1.289&[1.287,1.295]&0.011&[0.0039,0.044]&8&[1,8]&4&[1,10]\\
Iran&1.596&[1.563,1.709]&0.78&[0.71,0.79]&4&[3,4]&30&[29,40]\\
Ireland&1.401&[1.372,1.79]&0.083&[0.042,0.62]&2&[2,2]&20&[2,40]\\
Italy&1.242&[1.237,1.255]&3.88&[3.11,5.37]&19&[18,19]&6&[4,8]\\
Netherlands&1.387&[1.372,2.28]&0.34&[0.29,1.314]&4&[2,4]&35&[2,40]\\
Norway&1.461&[1.379,2.935]&0.164&[0.082,1.64]&1&[1,3]&20&[1,40]\\
Poland&1.476&[1.422,3.542]&0.109&[0.054,2.17]&1&[1,1]&20&[1,40]\\
Portugal&1.400&[1.356,3.077]&0.0886&[0.044,1.77]&1&[1,1]&20&[1,40]\\
South Africa&1.563&[1.513,2.65]&0.057&[0.031,1.15]&1&[1,2]&20&[1,37]\\
South Korea&1.238&[1.231,1.289]&0.98&[0.76,4.65]&20&[14,20]&7&[1,9]\\
Spain&1.411&[1.405,1.418]&0.127&[0.016,0.63]&20&[20,20]&5&[1,40]\\
Sweden&1.356&[1.343,1.372]&0.064&[0.0068,0.27]&19&[19,20]&3&[1,40]\\
Switzerland&1.527&[1.448,2.121]&0.4&[0.267,1.078]&3&[2,3]&12&[2,21]\\
UK&1.252&[1.248,1.257]&0.68&[0.412,0.878]&19&[17,20]&1&[1,1]\\
USA&1.306&[1.303,1.309]&0.0014&[0.0006,0.009]&10&[3,12]&4&[1,7]
\end{tabular}
}
\medskip
\caption{Fit parameters for 20 countries.\label{tab:app:country-fits}}
\end{table}

\subsection{Explaining \math{\beta}}

We perform a simple statistical analysis to test if
\math{\beta} can be explained by any of the country parameters
in Table~\ref{tab:app:country-comp}.
We include the delay as a global explanatory variable, which would
account for a global increase in vigilence as time passes and
awareness of the pandemic increases. One expects \math{\beta} to
decrease with the delay. A table of correlations of
\math{\beta} with the various parameters is shown below.
For our analysis we use the best case \math{\beta}, although similar
results follow from the optimal \math{\beta}.
\mandc{\text{
\begin{tabular}{r|ccccccc}
Dependent var.&Delay&Den&Age&Wealth&Income&Poverty&Den-Init\\\hline
  \math{\rho(\beta,x)}&0.68&-0.15&-0.51&-0.26&-0.23&0.403&0.52\\
\math{p}-value& 0.001&0.52&0.02& 0.26&0.32&0.082&0.017
\end{tabular}
 }}
As expected, there is a very significant correlation of
\math{\beta} with delay, but in the opposite direction.
\begin{itemize}
\item
  The larger the delay, the larger is \math{\beta}. The more a country has
  observed, the faster the spread in that country. That seems unusual but
  seems strongly indicated by the data.
\item Population density at the infection site has a strong
  positive effect but the country's population density does not.
\item
  There is faster spread in poorer countries.
\item
  Median age has a strong effect. Spread is faster in younger countries.
  The youth are more mobile and perhaps also more carefree.
\item There is a slight negative effect from wealth and per-capita income.
  Spread is slower in richer countries. Perhaps this is due to
  more risk-aversion, perhaps higher levels of education, perhaps less
  use of public transportation. Whatever the cause, it does have an
  impact, but relatively smaller than the other effects.
\end{itemize}
We now use regularized
regression to perform
a linear model fit to explain \math{\beta}.
To make the weight magnitudes meaningful, we normalize the data.
We use a
leave-one-out
cross validation to select the optimal regularization parameter
(which happens to be 20). The optimal regularized fit with this
regularization parameter gives a new feature
\mandc{
  X=
  w_1\cdot(\text{\small Del})+
  w_2\cdot(\text{\small Pop})+
  w_3\cdot(\text{\small Age})+
  w_4\cdot(\text{\small Wlth})+
  w_5\cdot(\text{\small Inc})+
  w_6\cdot(\text{\small Pov})+
  w_7\cdot(\text{\small Pop-Init})
}
The learned weights and the their ranges which yield a cross-validation error
within 10\% of optimal are shown in the table.
\begin{figure}[t]
    \centering
    \includegraphics[width=0.6\textwidth,trim={0cm 0cm 8.1cm 16.25cm},clip]{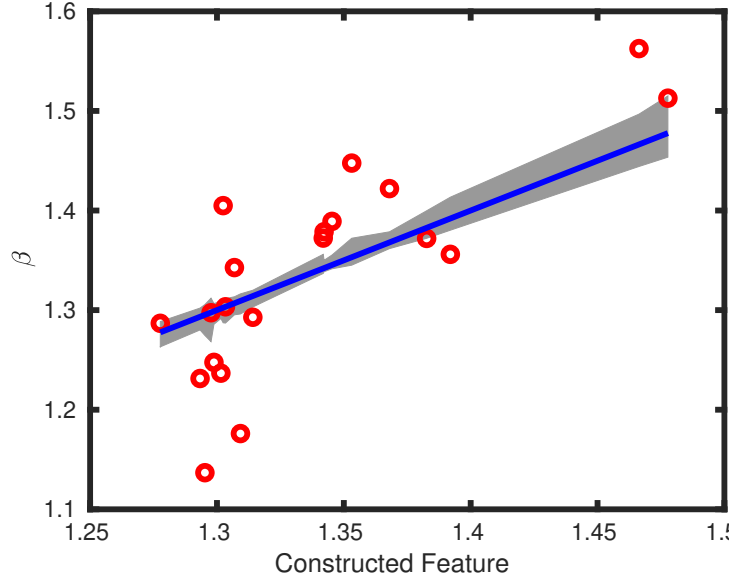}
    \caption{Optimal feature to predict \math{\beta}, within a
      cross-valudation setting to select the regularization parameter. The
      gray region is the range of the predicted value for each country.
      The \math{R^2=0.57}, so the cross-validation based optimal
      linear feature captures 57\% of the residual variance.
    \label{fig:app:betapred}}
\end{figure}
\mand{
  \scalebox{0.9}{\text{\tabcolsep15pt\renewcommand{\arraycolsep}{1.2}\begin{tabular}{l|lc}
Feature&Weight&\math{[\min,\max]}\\\hline
Delay&  0.030705 &       [0.023,0.045]\\
Pop-Den&   -0.0014362&   [-0.002,   0.00064]\\
 Age   &-0.012958    &[-0.013,    -0.011]\\
Wealth  &  -0.004016 & [-0.004,   -0.0028]\\
Income   &-0.0032445  & [-0.0033,   -0.0031]\\
  Poverty &    0.0109  & [0.0089,      0.016]\\
Pop-Den-Init&     0.018305  &  [0.015,     0.025]\\
\end{tabular}
}}}
The predictions of \math{\beta} using this feature are also shown in
Figure~\ref{fig:app:betapred}.
A statistical regression model using these data produces the fit:
\mandc{
\scalebox{0.9}{\text{\tabcolsep15pt\renewcommand{\arraycolsep}{1.2}
\begin{tabular}{r|cccc}
Feature&  Estimate &         SE   &\math{t}-Stat   &  \math{p}-Value  \\\hline
    (Intercept)&  1.2213 &      0.32759&      3.7282 &   0.0028836  \\
    Delay     &   0.0030891&     0.0010042&      3.0762&    0.0096064  \\
    Pop-Den  &   2.0281e-05&    0.00014317&     0.14166  &     0.8897  \\
     Age    &    -0.0013599  &   0.0074969  &   -0.1814&      0.85908  \\
    Wealth   &   2.274e-06&    7.9597e-06 &    0.28569 &     0.77999  \\
    Income  &    -1.8495e-06&    6.0964e-06 &   -0.30338&       0.7668  \\
    Poverty   &   0.0019366 &    0.0024657 &     0.7854 &     0.44745  \\
    Pop-Den-Init& 9.6516e-06&    6.6452e-06 &     1.4524 &     0.17203  \\
\end{tabular}
}}}
The statistical regression model also identifies positive weights on
delay, population density at the initial site and poverty in that order
of significance.

As we observed from the correlations, Delay, Poverty and Population Density
at the initial infection site have strong positive weights.
Age has a strong negative weight. Wealth and income have weak negative
effects, but non-zero. The population density of the country as a whole
seems to have no effect, with a weight range that includes 0.

\end{document}